\documentclass[11pt,twoside]{article}

\usepackage{asp2006}
\usepackage{epsf}
\usepackage{lscape}

\def\apjl{ApJL}
\def\apj{ApJ}
\def\apjs{ApJS}
\def\mnras{MNRAS}
\def\nat{Nature}
\def\aj{AJ}
\def\aap{A\&A}

\begin{document}
\title{Galaxy formation \& evolution: the far-ir/sub-mm view}   
\author{M. Cirasuolo(1) \& J. S. Dunlop(1,2)}   
\affil{ (1) Institute
for Astronomy, University of Edinburgh, 
Royal Observatory, Edinburgh, EH9 3HJ, UK.\\
(2) Department of  Physics and  Astronomy, University of British Columbia, 
6224 Agricultural Rd., Vancouver, B.C., V6T 1Z1, Canada.}

\begin{abstract} 
We review our current knowledge of the population of high-redshift 
sub-mm/mm galaxies, with particular emphasis on recent results from the 
SCUBA HAlf Degree Extragalactic Survey (SHADES). All available evidence 
indicates that these objects form the high-redshift, high-luminosity, 
high-mass tail of the dusty starforming galaxy population revealed 
at lower redshifts and luminosities by Spitzer. Current theoretical models
of galaxy formation struggle to reproduce these extreme objects in the 
numbers indicated by current surveys.

\end{abstract}

\section{Introduction}
It is now 10 years since the advent of array cameras on sub-mm/mm telescopes
provided the first discoveries of sub-mm/mm galaxies (SMGs; Smail et al. 
1997). Over the intervening decade a series of surveys
at 850$\mu$m 
with SCUBA
(Smail et al. 1997; Barger et al. 1998; Hughes et al. 1998; 
Eales et al. 1999, 2000; Cowie et al. 2002; Scott et al. 2002; Webb et al. 2003;
Serjeant et al. 2003; Wang et al. 2004; Coppin et al. 2006) 
and at millimeter wavelengths with
MAMBO and AzTEC 
(Greve et al. 2004; Dannerbauer et al. 2004; Carilli et al. 2005; 
Laurent et al. 2005; Bertoldi et al. 2007) have been undertaken in order to 
clarify the prevalence and properties of SMGs, and their role in the 
process of galaxy formation.

Multi-frequency follow-up has shown that SMGs 
are heavily dust-obscured galaxies at high redshift ($z > 1$) 
undergoing intense star-formation with inferred star-formation rates of 
several hundred to a few thousand solar masses per year.
These dusty star-forming galaxies have a comoving number density of 
$1-3 \times 10^{-5} \rm Mpc^{-3}$, approximately 
comparable to the number density 
of $2-3 L^*$ massive elliptical galaxies observed in the local Universe
(Scott et al. 2002). This, combined with preliminary measurements of their 
dynamical masses (Greve et al. 2005; Tacconi et al. 2006) 
and clustering properties (Scott et al. 2006), suggests that SMGs  
represent an important phase 
in the formation of present-day massive galaxies. However, SMGs have not proved
to be a natural prediction of semi-analytic models of galaxy formation
(Baugh et al. 2005). Current work in this field is thus focussed on providing
improved data on the SMG population of sufficient quality to allow detailed
statistical comparison with theory.

\section{SHADES} 
Taken together, all the early sub-mm surveys (Clusters - Smail et al. 1997; 
Hubble Deep Field - Hughes et al. 1998; Canada-UK Deep Sub-mm Survey - Eales et
al. 1999; Hawaii Flanking Fields survey - Barger et al. 1999; 8-mJy survey -
Scott et al. 2002; 8-mJy IRAM MAMBO follow-up - Greve et al. 2004;
HDF "supermap" - Borys et al. 2003)
cover $\simeq 500$ $\rm
arcmin^2$, but with widely varying depth and data quality. While 
interesting results have been extracted from a combined re-analysis of 
these surveys
(e.g. Scott et al. 2006), this early work demonstrated the need 
for a much larger, homogenous, flux-limited sub-mm/mm survey. In an attempt 
to fulfil this requirement, the SCUBA HAlf 
Degree Extragalactic Survey (SHADES) was commenced at the JCMT in 2003.

SHADES was designed to provide the first homogeneous sub-mm survey of 
sufficient size to provide a complete flux-limited sample of several hundred 
SMGs. The aim was to achieve a clean measurement of the number 
counts and redshift distribution of bright 
($\rm S_{850\mu m} \simeq 8$ mJy ) SMGs, along with a first measurement 
of the strength of their angular clustering (van Kampen et al. 2005; 
Mortier et al. 2005). The relatively shallow depth of the survey
($\sigma_{850} \simeq 2$ mJy) was chosen both to minimise the problems 
of source confusion (given the 14-arcsec JCMT beam
at 850$\mu m$), and to maximise
the effectiveness of multi-frequency follow-up. In addition, it is the 
brightest star-burst galaxies which appear to offer the 
most challenging test of 
theoretical models of galaxy formation (e.g. Baugh et al. 2005).

\begin{figure}
\plotone{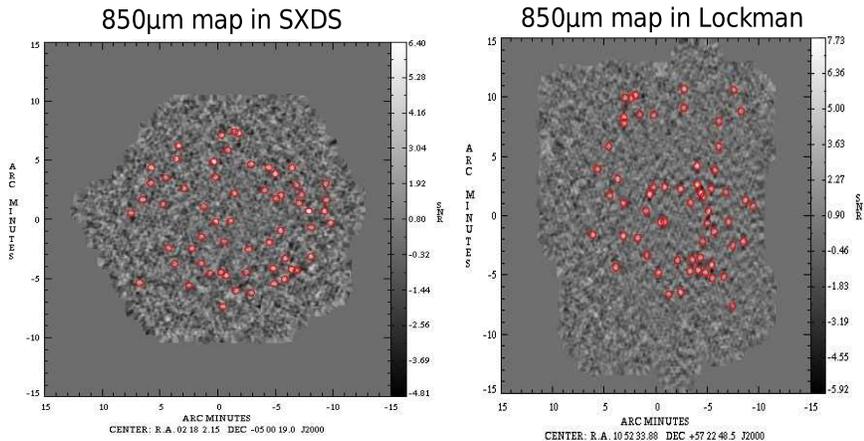}
\caption{\label{fig1} The SHADES 850$\mu$m S/N maps in the SXDS (left) and
Lockman Hole (right). The extracted sources are highlighted with red circles.}
\end{figure}

The SHADES 850$\mu$m maps were made over a period of 
3 years with an increasingly ailing SCUBA camera, eventually 
retired due to cryogenic problems. The final sub-mm data
cover an area $\sim 1/4$ of square degree down 
to an rms depth of $\sigma_{850} \simeq 2$mJy in two fields with deep 
multi-wavelength
ancillary data: the Lockman Hole and 
the Subaru/XMM deep field (SXDF). While short of the original goal, the 
final maps still 
form the largest extragalactic sub-mm imaging survey of meaningful
depth undertaken to date, and provide a uniquely powerful resource for the 
study of the bright sub-mm population
(Figure \ref{fig1}).

\begin{figure}
\plotfiddle{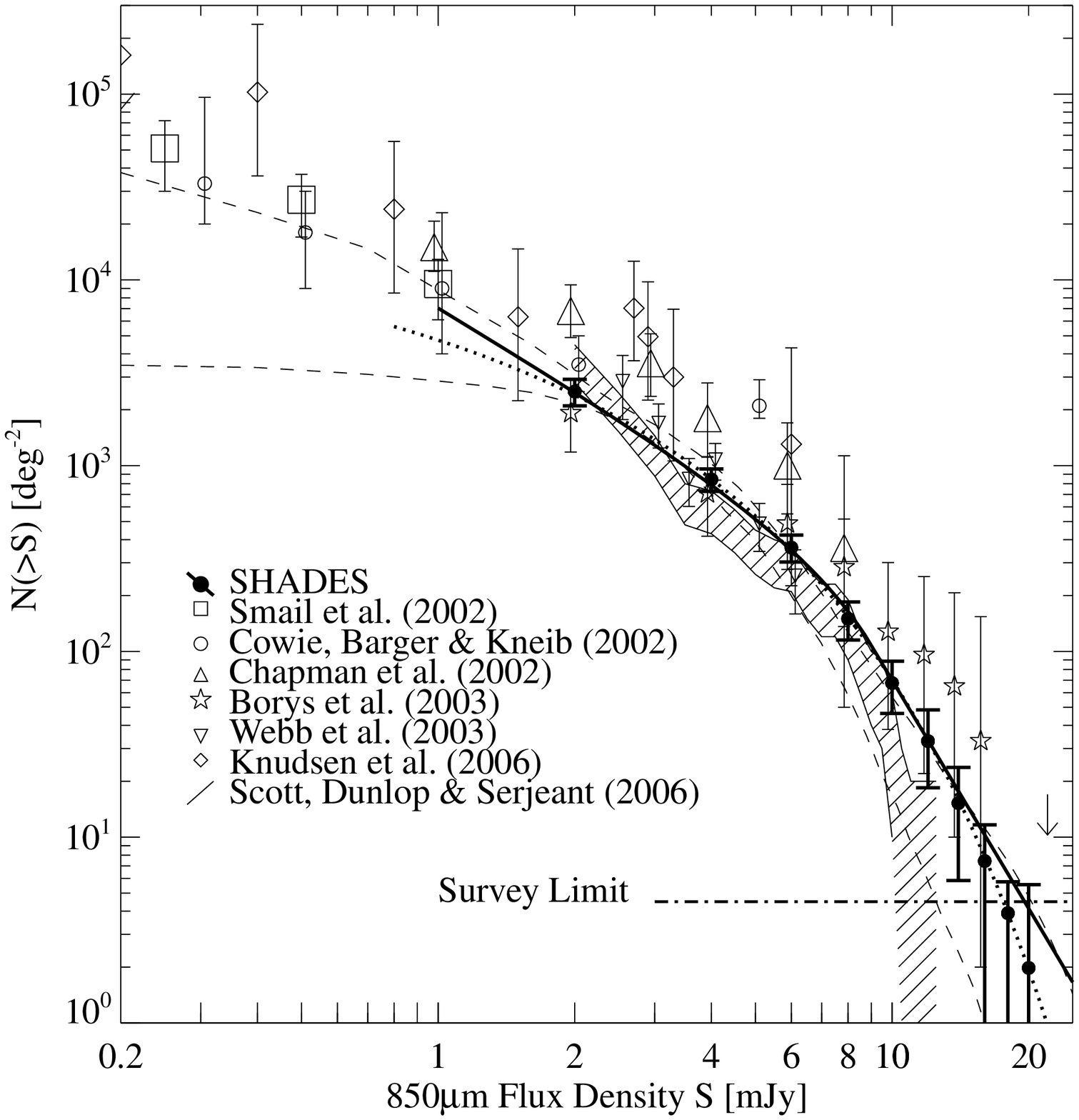}{9.5cm}{0}{40}{40}{-130cm}{0cm}
\vspace{-2cm}
\caption{\label{fig2} The 850$\mu$m number counts as derived from the 
SHADES SCUBA maps shown in Figure 1. The result is the best estimate to date
of the number counts in the 1-10 mJy range, and resolution of 20-30\% of the 
background.}
\end{figure}

The 850$\mu m$ data have been subjected to an extremely thorough analysis, 
with 4 independent sub-groups undertaking independent 
data reduction and source extraction. The final merged results of this process
are reported in Coppin et al. (2006), and include a new sample of 120 SMGs 
with statistically deboosted fluxes, and a definitive measurement of the 
850$\mu m$ source counts in the range $1 - 10$ mJy (Figure \ref{fig2}). 

\section{Identifications}
As a result of the rather poor positional accuracy (typically 2--3$''$
r.m.s.) provided by single-dish sub-mm/mm detections, further 
long-wavelength follow-up observations have proved essential for the 
robust identification of SMG galaxy counterparts.
An attractive and unambiguous 
way to refine sub-mm source positions is to obtain 
deep interferometric follow-up 
observations at mm/sub-mm wavelengths, as demonstrated 
by IRAM PdBI and SMA observations of some of the brightest SMGs discovered 
to date (Downes et al.\ 1999;
Dannerbauer et al.\ 2002, 2004, 2008; Younger et al.\ 2007; Wang et
al.\ 2007). However, current mm/sub-mm interferometry is still too demanding 
and expensive to be applied to large statistical samples of SMGs, and 
wide-field radio imaging continues to offer the best route for refining 
the positions of the majority of bright SMGs (Ivison et al.\ 2002).

\begin{figure}
\plotone{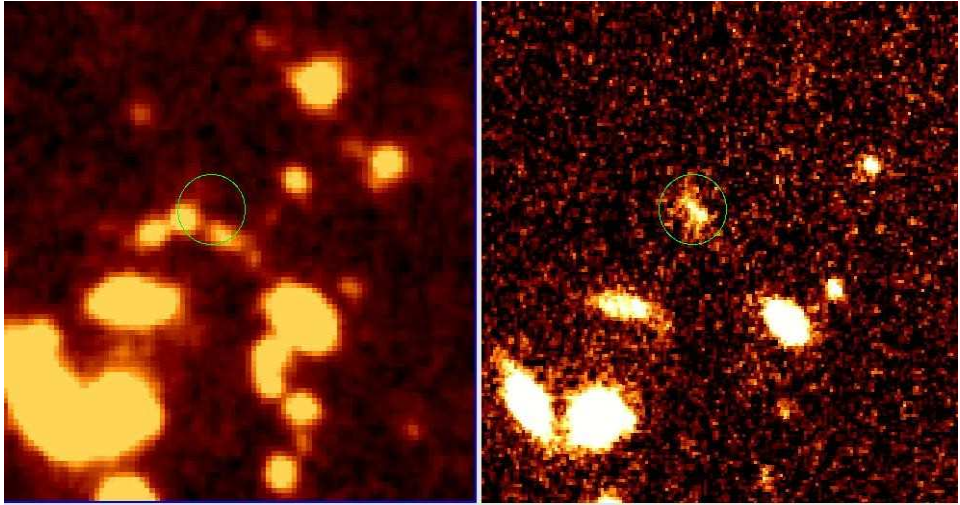}
\caption{\label{fig3} Optical and near-infrared images of the field around 
a sub-mm galaxy (Iono et al. in prep). 
In both images the exact position of the sub-mm galaxy as obtained
by SMA interferometry is indicated by the green circle.
The true galaxy counterpart
is clearly seen in the $K$-band, but completely invisible in the 
$R$-band.}
\end{figure}

Within the SHADES fields, deep VLA imaging at 1.4 GHz has allowed
$\simeq 70$\% of the SHADES sources to be identified  with radio counterparts,
yielding positions accurate to $\simeq 1$ arcsec (Ivison et al. 2007).
The vast majority of the resulting identifications are unique and unambiguous,
allowing the use of SMA follow-up to be focussed on those sources for which 
it is required to differentiate between alternative radio counterparts
(Younger et al. 2008). The importance of this
careful two-stage approach is nicely demonstrated by the case of a SHADES source 
which has three 
alternative statistically acceptable radio counterparts (Ivison et al.
2007). Deep SMA observations
(Iono et al. 2008, in prep) have now shown that the sub-mm emission is 
confined to one of the radio
counterparts, and is located at the position of a galaxy which is 
invisible in the deep Subaru optical images (to $R > 27$), but is clearly 
detected in the near-infrared (Figure \ref{fig3}).
The case of this sub-mm source demonstrates the power of sub-mm interferometry, but
also the importance of deep near-infrared data for an unambiguous
identification and study of the host galaxy. 

Finally, mid-infrared observations with Spitzer have also been used 
to further raise the identification fraction in SMG samples, including SHADES 
(Egami et al.\ 2004; Ivison et al.\ 2004; 2007). 
However, some of the latest SMA results demonstrate 
that apparent Spitzer identifications of radio-unidentified 
SMGs need to be treated with caution
(Younger et al. 2007; 2008).

\section{Redshifts}

The ability to locate the position of the SMGs to arcsec accuracy
has allowed deep spectroscopic observations of SMG counterparts
at both optical (Chapman 2003, 2005; Swinbank et al. 2004)
and more recently mid-infrared wavelengths
(Menendez-Castro et al.\ 2006; Valiante et al.\ 2007; Pope et al.\
2008). However, because the optical counterparts of SMGs are generally
very faint and red, the number of SMGs with robust spectroscopic redshifts
remains rather low (only 10\% of the SHADES sources currently have
unambiguous spectroscopic redshifts).

Given the difficulty (and in some cases impossibility) of determining 
optical spectroscopic redshifts for SMGs, a significant amount of effort has 
been invested in the development of redshift estimation techniques based 
on both radio/sub-mm and infrared/optical photometry.
The radio:sub-mm flux-density ratio was first proposed as a 
crude (but potentially unbiased and complete) method for estimating the 
redshifts of SMGs by Carilli \& Yun (1999). The use of long-wavelength 
photometry has been developed further by Aretxaga et al. (2003), and 
applied to the SHADES sample in Aretxaga et al. (2007). The results 
broadly confirm the findings of previous studies - the vast majority of the
bright SMGs have redshifts in the range $2 < z < 3$, with very few SMGs found at $z < 1$, and little evidence for a significant high-redshift tail beyond
$z \simeq 4$. Interestingly, it is generally the brightest SMGs which appear
to lie at the highest redshifts (e.g. Younger et al. 2008), providing 
further evidence of the existence of down-sizing in cosmic star-formation 
history.

More accurate redshift estimates can be obtained from the more extensive 
optical+near-infrared+Spitzer-IRAC photometry, albeit this is necessarily
confined to SMGs with clearly detected infrared counterparts. Within 
SHADES this technique has been applied by Dye et al. (2008) in the Lockman
Hole field, and by Clements et al. (2008) in the SXDF. These studies 
have also enabled an exploration of the stellar masses and star-formation
histories of the SHADES galaxies. The results indicate that bright sub-mm 
sources are housed in galaxies which are already massive (typically a few $\times 10^{11} {\rm M_{\odot}}$) with the ``current'' starburst having been 
preceded by previous star-formation events which formed at least half 
of the final stellar mass.

\section{The nature of sub-mm galaxies}
Armed with reliable identifications and redshift estimates 
for SMGs it is possible to begin to study 
the physical properties of these sources. In particular, to place 
SMGs in the broader context of galaxy evolution, it is important 
to investigate the masses, sizes and morphologies of the galaxies which 
host these most luminous dust-enshrouded starbursts.

Studies of the rest-frame optical/UV light of sub-mm sources 
have revealed a complex and often disturbed morphology (Smail et al. 2004;
Conselice et al. 2003; Chapman et al. 2004; Swinbank et al. 2006) 
suggesting major interactions and/or non
uniform dust obscuration. However, very little is known about the morphology at
longer wavelengths, which better traces the stellar mass of the system. 
In particular, it is still unclear whether sub-mm galaxies are disc
galaxies (as might be expected for gas-rich star-forming systems) or
massive ellipticals (as perhaps suggested by their number density,
which matches that of present-day $> 2 -3 L^*$ elliptical galaxies).

\begin{figure}
\plottwo{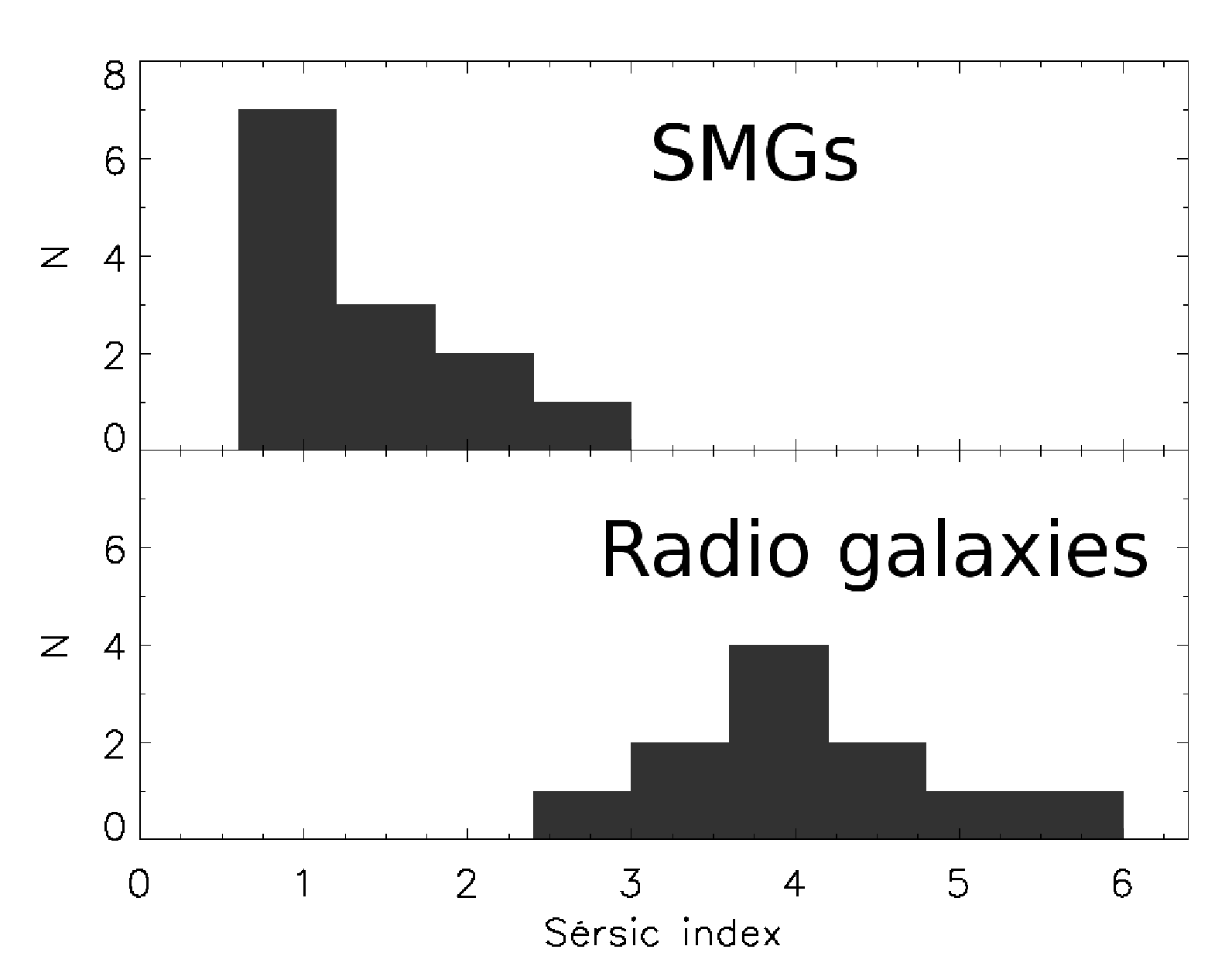}{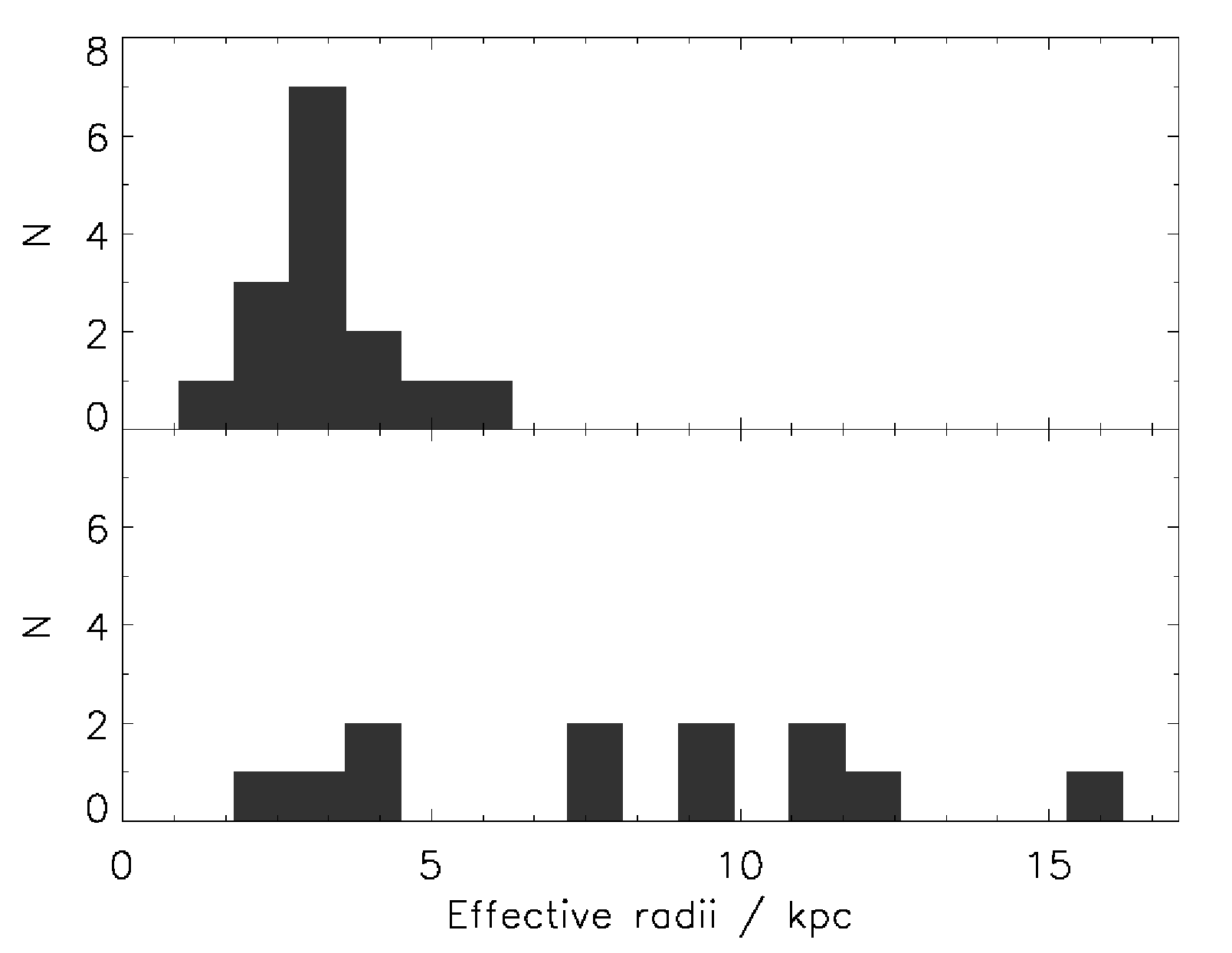}
\caption{\label{fig4} Histogram of the S\'{e}rsic index (left panels) and
effective radii (right panels) derived by using deep, high resolution $K$-band
imaging for 15 bright sub-mm galaxies at the top and radio galaxies
in the control sample at the bottom (Targett et al. 2008 in prep).}
\end{figure}

A limited number of sub-mm galaxies have been studied in the infrared
(e.g. Smail et al. 2004; Swinbank et al. 2006), but to date no systematic 
study of the morphologies of sub-mm galaxies has been undertaken. 
To judge the relevance of any size and morphological information
which might be gleaned from the imaging of the sub-mm
galaxies, it is also important to assemble data of comparable quality
for a well-defined control sample. Recently, Targett et al. (2008, in prep.) 
have obtained
deep, high resolution (0.5 arcsec) $K$-band imaging of 15 bright 
($\rm S_{850\mu m} >$8mJy) sub-mm
galaxies at $z \simeq 2$ together with a sample of 13 radio galaxies 
at the same redshift. This latter sample was chosen as a control 
because radio galaxies are the most massive galaxies in existence at this
epoch and will certainly evolve into  massive ellipticals.
As shown in Figure \ref{fig4}, SMGs are very compact with an effective radius
of $\simeq 3$ kpc, significantly smaller than the radio galaxies at the same
redshift. The analysis of the light profile also reveals a substantial
difference
between radio and sub-mm galaxies. The distribution of the 
S\'{e}rsic indexes shown
in figure \ref{fig4} suggests that the SMGs are mostly discs while the radio 
galaxies, as perhaps expected, are generally 
found to be de Vaucouleurs spheroids
(Targett et al. 2008, in prep).

These results are also in agreement with recent high-resolution 
molecular CO measurements of a sample of bright sub-mm galaxies at $z > 2$
obtained by Tacconi et al. (2006). These CO observations reveal that the 
millimeter line and continuum is compact with sizes $< 4$ kpc. A large 
fraction 
($\simeq$ 60\%) of the sources with CO detections show a double-peaked line 
profile (Tacconi et al. 2006; Neri et al. 2003; Greve et al. 2005), 
indicative of orbital motion. However, with the current resolution it is not 
possible to distinguish between a single rotating disc and two galaxies in a 
final stage of merging. It is also worth noting that recent deep, 
high-resolution radio observations of a sample of 12 sub-mm galaxies found 
rather extended radio emission with  
linear sizes in the range  $1-8$ kpc (Biggs \& Ivison 2008) 

Another parameter of fundamental importance for 
understanding the nature of
SMGs is galaxy mass.
Measurements of molecular gas mass via CO observations (Neri et al. 2003;
Greve et al. 2005) and dynamical masses (Tacconi et al. 2006;
see also Swinbank et al. 2006 for masses derived via integral field 
spectroscopy) all suggest that 
sub-mm galaxies are gas rich ($\rm M_{gas} \simeq 10^{10} -10^{11} M_{\odot}$)
and massive ($\rm M_{dyn} \simeq 5 \times 10^{10} M_{\odot}$).
These estimates are also in agreement (within the errors) with
stellar mass estimates ($\rm M_{stars} \ge 10^{11} M_{\odot}$) 
obtained by fitting the optical and near-infrared
spectral energy distribution (Borys et al. 2006;
Clements et al. 2008; Dye et al. 2008).

\section{Discussion and future prospects}
It is sometimes asserted that sub-mm galaxies are bizzare objects caught 
in a very unusual or extreme 
phase/mode of star formation. However, their observed properties in fact seem 
largely as expected on the basis of extrapolation from less extreme starburst
galaxies at lower redshift. For example, as shown by Bouch\'{e}
et al. (2007), because of their small sizes and large gas densities,
SMGs lie at the
high surface-density end of the apparently
universal ``Schmidt-Kennicutt'' relation. In other words, 
an object
with a mass of $10^{11} \rm M_{\odot}$ in gas within a radius of few $kpc$ 
is 
{\it expected} to produce $\sim$1000 $\rm M_{\odot}$ of stars per year. 
Similarly, it is also ``expected'' that such extreme starbursts 
should be hosted by some of the most massive galaxies in existence 
at $z > 2$, given the relationship between 
stellar mass and star-formation rate at 
$z \simeq 2$ found by Daddi et al. (2007).

In conclusion, sub-mm galaxies appear to be massive gas rich discs engaged 
in a major star-formation event triggered, in at least some cases, by a
``recent'' galaxy-galaxy interaction. Given the starburst is relatively 
compact, one can speculate that we are witnessing these galaxies completing
the formation of their cores. Although the sub-mm galaxies are best described 
as discs, in terms of density (and space density) 
they are much more like ellipticals/bulges than present day star-forming 
spiral galaxies. 
The SMGs we observe at high redshift thus seem destined to evolve into massive,
passive spheroids, awaiting relaxation and further extended mass growth 
(e.g. by dry mergers) by a factor $\simeq$ two.

Over the next 3-5 years we can expect our understanding of SMGs to be 
placed on a much firmer footing be a series of genuinely revolutionary 
new projects. First, at the time of writing the SHADES consortium is completing
a new, expanded study of the SHADES fields using 1.1mm maps of the full
0.5 square degree area made with AzTEC on the JCMT (Austermann et al. 2008, 
in prep). Second, SCUBA2 is currently being installed at the JCMT in Hawaii,
and is expected to commence the SCUBA2 Cosmology Legacy Survey in early 2009.
Third, Herschel is due to launch later this year, and will provide complete
SEDs and bolometric luminosities for large numbers of the SMGs uncovered 
with SCUBA2. Finally, these photometric surveys will set the stage for the
detailed astrophysical study of SMGs with the Atacama Large Millimetre 
Array.

\acknowledgements 
The authors gratefully acknowledge the many contributions 
of SHADES consortium members to the work described here.

\end{document}